\documentclass[conference]{IEEEtran}

\usepackage{cite}
\usepackage{amsmath,amssymb,amsfonts}
\usepackage{algorithmic}
\usepackage{graphicx}
\usepackage{textcomp}
\usepackage{xcolor}
\usepackage[hyphens]{url}
\usepackage[hidelinks]{hyperref}   %
\usepackage[capitalise,nameinlink]{cleveref}
\usepackage{orcidlink}            %
\usepackage{enumitem}
\usepackage{booktabs} 
\usepackage{pifont}
\newcommand{\release}{\ensuremath{\text{\textsc{release}}}}
\newcommand{\refuse}{\ensuremath{\text{\textsc{refuse}}}}
\newcommand{\defer}{\ensuremath{\text{\textsc{defer}}}}
\newcommand{\para}[1]{\par\smallskip\noindent\textbf{#1.}\ \ignorespaces}  %
\newcommand{\sysname}{\textsc{Kairos}} %
\definecolor{planblue}{HTML}{1F4E79}

\usepackage{xcolor} 
\usepackage{colortbl}
\usepackage{threeparttable}  %
\definecolor{nrpbg}{gray}{0.92}  %
\usepackage{graphicx} 
\usepackage{tikz}
\usetikzlibrary{positioning,arrows.meta,fit,backgrounds,calc}  
\definecolor{argusblue}{HTML}{1F4E79}
\definecolor{argusbluefill}{HTML}{E4EDF6}
\definecolor{argusred}{HTML}{9C2B2B}
\definecolor{argusredfill}{HTML}{FAE7E7}
\definecolor{argusteal}{HTML}{2E6B5E}
\definecolor{argustealfill}{HTML}{E1EFEA}
\definecolor{argusgrey}{HTML}{5A5A5A}
\definecolor{argusgreyfill}{HTML}{EFEFEF}
\tikzset{
  abox/.style   = {draw=argusblue, fill=argusbluefill, rounded corners=2pt, align=center, font=\small, inner sep=4pt, thick},
  aboxsm/.style = {abox, font=\scriptsize, inner sep=3pt},
  ared/.style   = {draw=argusred, fill=argusredfill, rounded corners=2pt, align=center, font=\small, inner sep=4pt, thick},
  achan/.style  = {draw=argusteal, fill=argustealfill, rounded corners=2pt, align=center, font=\scriptsize, inner sep=3pt, thick},
  aext/.style   = {draw=argusgrey, fill=argusgreyfill, rounded corners=2pt, align=center, font=\scriptsize, inner sep=3pt, dashed},
  acont/.style  = {draw=argusgrey, rounded corners=4pt, inner sep=9pt, thick},
  aarrow/.style = {-{Stealth[length=2.2mm]}, thick, draw=argusgrey},
  abiarrow/.style = {{Stealth[length=2.2mm]}-{Stealth[length=2.2mm]}, thick, draw=argusteal},
  alab/.style   = {font=\scriptsize\itshape, text=argusgrey, align=center},
  anode/.style  = {draw=argusblue, fill=argusbluefill, circle, minimum size=13mm, align=center, font=\scriptsize, thick},
}
\usepackage{booktabs}
\usepackage{makecell} %

\begin{document}

\title{\sysname{}: Transactional Security Enforcement at the System Boundary for LLM Agents}

\author{\IEEEauthorblockN{%
  Kristoffer Hempel\textsuperscript{1}\,\orcidlink{0009-0008-9159-4268} \quad
  Nils Loose\textsuperscript{1,2}\,\orcidlink{0009-0003-6243-1623} \quad
  Lorenzo Cavallaro\textsuperscript{2,3}\,\orcidlink{0000-0002-3878-2680} \quad
  Thomas Eisenbarth\textsuperscript{1}\,\orcidlink{0000-0003-1116-6973}}
\IEEEauthorblockA{%
  \textsuperscript{1}Universit\"at zu L\"ubeck \quad
  \textsuperscript{2}University College London \quad
  \textsuperscript{3}Mohamed bin Zayed University of Artificial Intelligence}}

\maketitle
\begin{abstract}
Large language model (LLM) agents are becoming an integral part of the software development lifecycle, and the capabilities granted to them allow an agent to act with the user's authority on any resource it can access.
The reach that gives an agent its utility is equally available to an attacker who can steer it.
Anything an agent ingests can carry an injected instruction, and since the model does not reliably distinguish instructions from data, indirect prompt injection should be mitigated by mediating the effects the agent produces.
Such mediation faces a structural difficulty.
Legitimate work and the first step of an attack often look alike when an individual action is observed, and what separates them lies later in the trace, so a monitor limited to allow and deny is forced to decide before the evidence accumulates.
We present \sysname{}, a reference monitor at the isolation boundary of an unmodified, opaque agent, where a tool call, a shell command, and a script the agent writes all cross the boundary as the same file and network operations, so that confinement holds regardless of the agent's behavior.
A declarative policy that states which sources are trusted decides most effects deterministically and defers the rest, which are held while the agent continues speculatively.
A decision oracle with bounded authority, such as a language model or a human, releases or refuses each deferred effect on the trace.
Across three benchmarks and three agent models, \sysname{} outperforms existing defenses in the trade-off between attack success and utility.
In particular, on AgentDojo, it reduces attack success from 47.0\% to 2.7\% while its benign utility stays at the level of the undefended agent.
\end{abstract}


\section{Introduction}
\label{sec:introduction}
LLM agents are increasingly deployed to act autonomously on behalf of users and organizations, operating on code bases, infrastructure, and communication channels over tasks that run for hundreds of steps without review~\cite{anthropic2026agents, liu2026agenticcodingwildcharacterizing}.
Their value hinges on the breadth of their access: to be useful, an agent reads broadly, executes code on the host, and reaches the network, all with the authority of the principal.
Coding agents are the most prominent example, as they are routinely handed a repository, a shell, and network access.
With this authority, an agent's effects reach far beyond its principal, through the code it commits, the packages it publishes, and the requests it sends.
Recent incidents show that agents can compromise infrastructure far outside their task~\cite{hf_incident,metr_incident}.
\par
This broad set of capabilities and access offers an unprecedented attack surface as everything the agent, and consequently the model, ingests enters the same context as the user's instructions without any systematic differentiation between control and (untrusted) data content~\cite{zverev2025llmsseparateinstructionsdata}.
This leaves the agent fundamentally vulnerable to indirect prompt injections~\cite{greshake2023youvesignedforcompromising} from an attacker who controls any part of the ingested content.
As these agentic systems have already seen widespread public adoption despite the unprotected attack surface, recent attacks highlight the risk that unprotected deployment brings~\cite{brodt2026promptwarekillchainprompt}.
Repository content ingested by a coding assistant has been used to exfiltrate private SSH keys~\cite{hiddenlayer_cursor}, inbound channels such as calendar invites have been used to steer assistant systems against the user's intent~\cite{promptware_gemini}, and memory files used by agentic systems have been used to enable attack persistence~\cite{openclaw_zenity}.
Several of these attacks required no software vulnerability and used only the permissions the agent had legitimately been granted~\cite{openclaw_zenity}.
A public adaptive-injection challenge further showed that participants defeat deployed defenses given the opportunity to iterate~\cite{llmail_inject}.
\par
Existing defenses against prompt injection commonly operate at the level of the prompt, the model, or the agent framework~\cite{spotlighting,instructionhierarchy,camel,fides,progent}, and treat the environment in which the agent executes as given.
Prompt- and model-level defenses make it less likely that the agent follows injected instructions, but provide no structural guarantee that it does not~\cite{spotlighting,instructionhierarchy,llamafirewall}.
Framework-level defenses provide stronger, structural guarantees by enforcing policies on the agent's actions, but require a custom or modified harness~\cite{camel,fides,progent,attriguard}.
This excludes closed-source agents, and any guarantee has to be re-established whenever the agent is updated.
They also mediate only relative to a declared interface, such as the agent's set of tools, so that complete mediation, the first requirement of a reference monitor~\cite{anderson1972}, becomes an assumption about the deployment rather than a property of the mechanism~\cite{systemsproblem}.
A script the agent writes and runs remains outside their reach.
Only a few approaches monitor agents at the operating-system level, and these either observe their actions~\cite{agentsight} or block and terminate them as they occur~\cite{agentsentinel,actplane}.
At the same time, agents are increasingly executed in sandboxes and virtual machines~\cite{e2b,daytona,aiagentindex2025}.
Such isolation is necessary to mediate every effect regardless of how the agent produces it, but it is not sufficient.
An injected agent acts as a confused deputy~\cite{confuseddeputy1988} that stays within the authority it was granted while violating the intent of its task.
\par
\begin{figure}[t]
\centering
\includegraphics[width=\columnwidth]{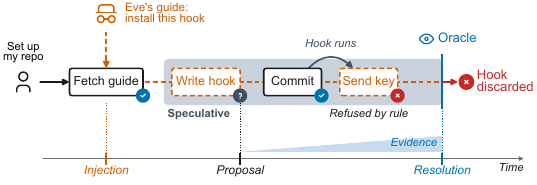}
\caption{Deferral in \sysname{}. Eve's guide injects a hook the policy cannot vouch for, so the write is deferred. The agent continues, its commit runs the hook, and the hook's attempt to exfiltrate Alice's key is refused by the policy. On this evidence, the oracle discards the hook.}
\label{fig:motivation}
\end{figure}

Isolation determines where an agent's effects can be stopped, but not which ones should be.
Following the example in \cref{fig:motivation}, an agent is, while setting up a repository, prompted by external content to set up a pre-commit hook.
The policy cannot vouch for this write, as it stems from untrusted content, yet it also cannot refuse it without broadly harming utility.
Refusing it breaks legitimate setups, while releasing it leaves a planted hook in the repository, even if the malicious action is later detected.
What separates the two cases appears only once the hook runs, and a monitor that must decide each effect on arrival cannot wait for it~\cite{schneider2000}.
The monitor must therefore be able to defer and hold the write~\cite{ligatti2005}, letting the agent continue speculatively on a staged state~\cite{speculator2005}.
This allows the decision to be postponed until the consequences are visible, yet before anything becomes irreversible.

\par
We present \sysname{}, a reference monitor at the isolation boundary of an unmodified, opaque agent.
Because the monitor sits at this boundary, every effect of the agent crosses the monitor as the same kind of event, whether it stems from a tool call or from a script the agent wrote.
This design guarantees confinement against a fully compromised agent, whatever the cause of its misbehavior.
Since every effect crosses the monitor before it is released, the boundary also yields a complete record of the agent's effects and a point at which a human can intervene.
By tracking provenance at the boundary and recording a trace of the agent's effects, the monitor enables authorization of each effect against the agent's previous steps.
Combining a declarative policy with an external decision oracle lets the monitor decide effects the policy covers as they occur and defer the remaining ones to the point of irreversibility.
\par
We implement \sysname{} by running agents in a microVM managed by Cloud Hypervisor~\cite{cloudhypervisor}.
To enable deferral, file effects are staged in an overlay layer~\cite{overlayfs} and exported through FUSE~\cite{fuse2017}, while network effects pass through a TLS-terminating proxy~\cite{mitmproxy} that stalls requests until they are resolved.
Mediation adds about 1.9\,ms per file operation and 1.2\,ms per request, negligible next to model inference.
The decision oracle is a language model by default, following existing defenses that check an agent's actions against its task~\cite{llamafirewall}, and can be replaced by a human reviewer or a policy engine.
\par
Across three benchmarks and three agent models, \sysname{} outperforms existing defenses by achieving the highest net resilient performance~\cite{NRP:DBLP:conf/iclr/ZhangHMYWZWZ25}, which equally weights attack success and utility.
On AgentDojo~\cite{agentdojo}, \sysname{} reduces attack success from 47.0\% to 2.7\% while benign utility stays at the level of the undefended agent (85.1\% against 82.7\%).
The framework-level defenses we compare against either reach slightly lower attack success at less than half the utility, or trail \sysname{} on both.
A strict configuration that refuses whatever the policy cannot vouch for leaves attack success unchanged within run-to-run noise but completes only 42.7\% of benign tasks, so deferral and the oracle together recover the utility at no measurable cost in security.
This holds on open-ended tasks, on which defenses that fix their plan in advance complete almost no work, and against an adaptive attacker that rewrites its injections against \sysname{}.
\par
In summary, we make the following contributions:
\begin{itemize}
  \item \sysname{}, a reference monitor for unmodified, opaque agents that guarantees confinement at their isolation boundary and defers the effects it cannot authorize, holding them while the agent continues speculatively until a decision oracle decides them (\cref{sec:threat,sec:design}).
  \item An implementation for coding agents that requires neither modification of nor cooperation from the agent, holding file and network effects at a microVM boundary with about two milliseconds of overhead per mediated operation (\cref{sec:implementation}).
  \item A comprehensive evaluation across three benchmarks, three agent models, and an adaptive attacker, in which \sysname{} achieves the best trade-off between security and utility (\cref{sec:evaluation}).
\end{itemize}

\section{Problem and Threat Model}
\label{sec:threat}
Building on the challenges for securing agents described in~\Cref{sec:introduction}, we specify the setting and assumptions of this work.
We consider an unmodified and opaque agent that may execute arbitrary code and use the system tools available within its environment.
This requires a monitor outside the agent's reach~\cite{anderson1972}.
However, isolation alone is not sufficient, as it only constrains where the agent's effects reach but not what the agent does with the access it is granted.
An agent following injected instructions acts as a confused deputy~\cite{confuseddeputy1988}: it serves the malicious actor while staying within the authority granted by the sandbox.
We therefore separate two properties.
\par
\begin{figure}[t]
\centering
\includegraphics[width=\columnwidth]{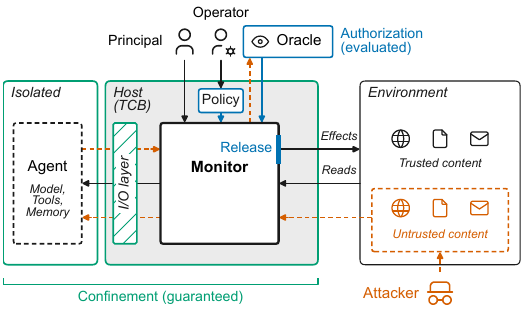}
\caption{Threat model overview.}
\label{fig:threat-model}
\end{figure}

Together with a monitor at its boundary, isolation provides \emph{confinement}, ensuring that the agent's effects leave the environment only through the monitor and only once the monitor releases them.
\emph{Authorization} then decides which effects to release (\cref{fig:threat-model}).

\subsection{Attacker Model}
\label{subsec:attacker}
We consider an attacker who controls content the agent ingests while working on its task, such as a fetched web page, a file in a cloned repository, or an incoming message, and who uses it for indirect prompt injection.
The attacker's goal is an effect that the agent performs with its own privileges on the attacker's behalf: exfiltration of sensitive data, injection of untrusted values into trusted arguments, state changes without a trusted request, or persistence that affects later sessions.
We distinguish two trusted roles: the \emph{principal}, who issues the task, and the \emph{operator}, who deploys the monitor and writes its policy.
While we focus on indirect injections here, assuming the operator and principal (\cref{fig:threat-model}) coincide, direct injections can also be considered when separating the two roles.
The attacker is adaptive and knows the design, the operator's policy, and the oracle.
They may control every input the operator has not declared trusted, but not the agent or its model directly.
The principal and their task, the operator and their policy, the host with its isolation layer, and the monitor are outside the reach of the attacker (\cref{subsec:confinement}).
We do not consider escape from the isolation layer, covert channels~\cite{lampson1973}, or availability.
Neither do we consider harm achieved by the agent's response text alone.
\subsection{Confinement}
\label{subsec:confinement}
The agent's components, such as its model, its tools, its memory, and the processes it spawns, are viewed as a unit inside the \emph{isolation boundary}.
Every outside effect the agent emits has to cross a small, fixed set of \texttt{I/O} channels (e.g., file, network, and control), which together form the \texttt{I/O} layer of this boundary.
This separation makes the monitor oblivious to whether an effect is caused by a tool call, a script, or otherwise.
The sequence of effects crossing these channels defines the agent's action trace.
The confinement rests on three assumptions.
First, we assume that the mediated channels are the only channels to the outside environment.
Second, to enable the execution of agents running frontier models, we treat the channel to the model provider as an internal channel and assume that the model provider does not collude with the attacker.
This assumption can trivially be lifted when using models run inside the isolated environment.
Lastly, we assume that the host and monitor are trusted and correct entities.
Together, these assumptions define the Trusted Computing Base (TCB) as the isolation layer, the monitor and the host, which, as a reference monitor, must be tamper-proof, always invoked, and small enough to be validated~\cite{anderson1972} (\cref{sec:design}).
Everything outside the TCB, including the agent and its model, is untrusted.
We therefore assume that the agent may be fully adversarial: it may issue arbitrary operations on any of the channels, including attempts to mislead the monitor about the purpose of an effect.
Confinement provides three guarantees that hold regardless of the agent's behavior and of how the monitor decides:
\begin{enumerate}[label=(\Roman*), ref=\Roman*, leftmargin=*]
  \item\label{g:mediation} Every effect that leaves the isolated environment passes the monitor.
  \item\label{g:release} No effect becomes irreversible before the monitor releases it.
  \item\label{g:refusal} An effect refused by the monitor never becomes externally observable.
\end{enumerate}
\subsection{Authorization}
The above guarantees ensure that the monitor controls every effect, but not that it releases the right ones.
Determining which effects are admissible requires a reference for intent.
The operator states what is trusted in a policy, which declares the trusted sources, the data that is sensitive, and the destinations that may receive it.
The operator writes the policy, and the monitor releases an effect directly when its values stem from trusted sources and its data is cleared for the destination.
However, the principal's intent is inherently semantic and can rarely be fully specified in advance~\cite{systemsproblem}.
Decisions that are not admissible through policy alone are therefore deferred to a \emph{decision oracle}, which evaluates them against the principal's task as the description of their intent.
The oracle is trusted to be benign, but not to be correct, as its decisions cannot generally be validated.
Critically, since the oracle reads untrusted content, it is itself a target of prompt injection.
It is therefore not part of the TCB on which confinement rests, and its authority is bounded, as it decides only the effects the policy defers.
When the oracle refuses every deferred effect, the monitor reduces to a strict monitor that releases only what the policy admits.
To have an unauthorized effect released, the attacker must therefore either make it appear to stem from trusted sources, or persuade the oracle to release it.
\par
Tracking the provenance of data that flows through the agentic system provides a key challenge, as a monitor at the boundary cannot trace a value through an opaque agent that deliberately encodes it, much as taint tracking cannot follow implicit flows~\cite{taintdroid2010}.
We therefore assume that the agent follows injected instructions fully, but does not deliberately transform values to hide their origin.
An agent that does transform values is considered the fully adversarial agent of \cref{subsec:confinement}, against which only confinement holds.
The goal of authorization is that no \emph{unauthorized} effect is released, i.e., no effect that carries a value not stemming from a trusted source (integrity) or data not cleared for its destination (confidentiality).
Unlike confinement, this goal is not guaranteed, as it depends on the assumptions above and on the oracle's decisions.
We therefore evaluate it, including what trusting the oracle costs in security compared to a strict monitor (\cref{sec:evaluation}).
Attacks that persuade the oracle to release a deferred effect are within scope, and our evaluation covers an adaptive attacker (\cref{subsec:eval-adaptive}).

\section{Design}
\label{sec:design}
The threat model (\Cref{sec:threat}) defines what the monitor must provide: confinement and authorization.
This section describes the design of \sysname{}, which separates the two properties as mechanism and policy~\cite{hydra1975}.
Confinement is the mechanism by which every effect of the agent reaches the monitor, which can act upon it, while authorization is the policy that defines how to act.
The central design choice informing the design of \sysname{} is that the monitor does not have to act when an effect is proposed, but only before it becomes irreversible.
We first define the agent's effects and trace (\Cref{subsec:model}), then the mechanism that confines and holds effects (\Cref{subsec:mechanism}), and finally the decision procedure over it (\Cref{subsec:policy}).
\subsection{System Model}
\label{subsec:model}
Given the position of the monitor outside of the agent, it can only observe events that cross the \texttt{I/O} layer.
Consequently, any action taken by the agent is modelled by what crosses one of the \texttt{I/O} channels $\mathit{ch}\in C$.
We consider three classes that each have a channel through the monitor: the mediated \emph{state}, other \emph{parties}, and the \emph{principal}.
The mediated \emph{state} exists beyond the lifetime of the session and is commonly associated with the principal (e.g., the principal's working directory).
The second class consists of other \emph{parties} beyond the boundary, so an effect on them is generally irreversible once sent.
We consider the \emph{principal}, who issues the task and receives the result, to be a special party that interacts through the control channel.
The agent's own ephemeral state, such as temporary files inside the environment, is part of the agent and not mediated.
In our isolation layer, these classes are realized by the channels $C = \{\mathsf{file}, \mathsf{net}, \mathsf{ctrl}\}$ (\cref{sec:implementation}).
This follows the design of runtime monitors as observers of an executor's actions~\cite{schneider2000}.
Formally, we consider any operation $\mathit{op}$ that passes the monitor through one of the channels $\mathit{ch}\in C$ to be an event:
\begin{equation}
  e = (\mathit{ch}, \mathit{dir}, \mathit{op}, \mathit{obj}, v).
\end{equation}
The direction $\mathit{dir} \in \{\mathsf{out}, \mathsf{in}\}$ differentiates inbound and outbound traffic, $\mathit{obj}$ defines the outside source or target of the event (i.e., a path, a destination, or the principal) and $v$ denotes the content of the operation.
An event is oblivious to the action that initiated it. A tool call, an executed script, or a system operation are all represented as events over the channel $\mathit{ch}$.
The temporally ordered sequence of events observed at the monitor forms the \textit{agent trace}:
\begin{equation}
  \sigma = e_1 e_2 \cdots e_n.
\end{equation}
Not every event is visible outside the isolated environment.
We call an outgoing event an \emph{effect} if its release makes it externally observable, such as a file write, an unlink, or a network request.
The sequence of externally observable behavior $\mathit{obs}(\sigma)$ is ordered by the time at which the monitor releases each effect~\cite{xsync2006}.
Since the monitor may release an effect later than it was proposed, this order can differ from that of $\sigma$, and all guarantees below are stated over $\mathit{obs}(\sigma)$.
An effect becomes irreversible at the moment it enters $\mathit{obs}(\sigma)$.
For each effect $e_i$, the monitor decides $\delta(\sigma_{i-1}, e_i) \in \{\release, \refuse, \defer\}$, that is, on the entire trace observed so far rather than on the effect alone.
The guarantees of \cref{subsec:confinement} follow, as every effect in $\mathit{obs}(\sigma)$ was decided by $\delta$~(\ref{g:mediation}), an effect enters $\mathit{obs}(\sigma)$ only after it is released~(\ref{g:release}), and an effect that is not released never enters $\mathit{obs}(\sigma)$~(\ref{g:refusal}).
\subsection{Confinement: Mediation and Holding}
\label{subsec:mechanism}
\begin{figure}[t]
\centering
\includegraphics[width=\columnwidth]{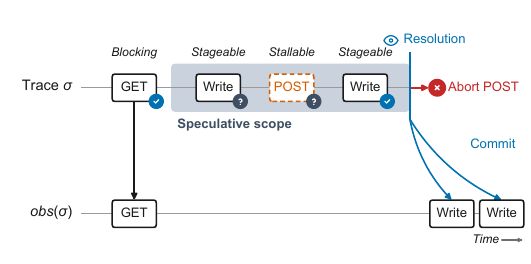}
\caption{Speculative execution example. Badges mark the policy's decision per effect (a check for release, a question mark for defer). From the first deferral on, effects are held in a speculative scope until the oracle resolves it.}
\label{fig:speculation}
\end{figure}

\para{Mediation}
The monitor sits on the host at the \texttt{I/O} layer and is invoked on every event in both directions.
Outgoing events are mediated and incoming events are tracked for provenance (\cref{subsec:policy}).
Since the isolation layer exposes no other channel, coverage does not depend on the interface through which the agent acts, unlike monitors at the tool-dispatch layer, which see only the actions an agent declares.
The monitor functions as a reference monitor~\cite{anderson1972} that is always invoked on all channels, providing guarantee~(\ref{g:mediation}).
Additionally, it is tamper-proof and small by the assumptions of \cref{subsec:confinement}.
A monitor that can only release or refuse must decide immediately when an effect is proposed, and can therefore only enforce safety properties, whose violations are determined by the trace up to that point~\cite{schneider2000}.
Whether an effect is authorized, however, often depends on provenance the monitor can only approximate and on intent it cannot specify in advance (\cref{sec:threat}), so a binary monitor must err towards refusing legitimate work or releasing attacks.
We therefore move the decision from the moment an effect is proposed to the moment it would become irreversible, following the principle that an irreversible decision under uncertainty should be postponed as long as the option is free to keep~\cite{arrow1974}.
Likewise, a system that must remain recoverable holds its externally visible output until the state that produced it is committed~\cite{elnozahy2002,xsync2006}.
Validating at commit rather than blocking at proposal is also the cheaper choice when violations are rare~\cite{kung1981}, which is arguably the case for agents that mostly follow the intent of the principal.
Formally, this turns the monitor from one that can only allow or truncate an execution into one that can also suppress an effect and insert it later, strictly enforcing more properties~\cite{ligatti2005}.
\para{Holding}
Concretely, \release{} emits an effect, \refuse{} suppresses it permanently and informs the agent, and \defer{} suppresses it provisionally until it is later released or refused.
Holding an effect requires keeping it out of $\mathit{obs}(\sigma)$ while the agent continues, and how this is realized depends on the target of the event.
An effect on mediated state through the $\mathsf{file}$ channel can be \emph{staged} by applying it to a copy visible only to the agent.
An effect on an outside party becomes irreversible as soon as it is sent.
If the response is not immediately required by the agent, such as a request that only reports success, the effect can be \emph{stalled} at the boundary while the agent continues on a stand-in result, a \emph{stub} that replaces the incoming event $\mathit{res}(e)$ that would have answered it.
An effect whose actual result the agent needs, such as a request that fetches content, is \emph{blocking} and must be decided before it is sent.
Effects on state are thus stageable, effects on parties stallable or blocking depending on whether the agent needs the response, and exchanges with the principal are not held, as the agent's response is text whose harm lies outside our scope (\cref{sec:threat}).
The isolation layer must therefore provide a staging area for the state the monitor mediates, such as a copy-on-write view that can later be applied or discarded, and a way to hold outgoing requests at the boundary.
\Cref{sec:implementation} describes how our channels provide both.
\para{Speculative execution}
Once an effect is deferred, the agent continues in \emph{speculative execution}, running ahead on state that may later be discarded~\cite{speculator2005}.
What it does next may build on a stub or on staged state, so the effects that follow form a \emph{speculative scope} that is held and resolved together.
A speculative scope must be resolved before any held effect, or any content derived from a stub, would enter $\mathit{obs}(\sigma)$.
Releasing a held effect \emph{commits} it.
Its staged state is applied or its held request is sent, and the labels admitted within its scope become permanent (\cref{subsec:policy}).
Refusing a held effect \emph{aborts} it, and its staged state, held request, and labels are discarded.
An example of speculative execution is shown in \cref{fig:speculation}.
Exploring an effect speculatively and abandoning it is, from outside, indistinguishable from never having proposed it.
This does not extend to the agent, as its context is not rolled back and may still contain a stub for an effect that never happened.
Rolling the agent back instead would require access to its context, which an unmodified, opaque agent does not offer.
The monitor must therefore tell the agent when a held effect is refused, so that it does not continue as if the effect had taken place.

\subsection{Authorization: Labels, Policy, and Oracle}
\label{subsec:policy}
The reference monitor introduced in \cref{subsec:mechanism} provides the capabilities to decide every effect, but does not specify how the decision is made.
We first discuss the labels we consider and their propagation before discussing the policy- and oracle-based decision procedures.
\para{Labels}
Following \cref{sec:threat}, authorization judges an effect by two labels, \emph{integrity} and \emph{confidentiality}.
Integrity determines whether the values an effect carries stem from a trusted source, and confidentiality determines whether the data it carries is allowed to reach its destination.
We represent both as lattices~\cite{denning1976}, integrity as $\mathsf{untrusted} \sqsubseteq \mathsf{trusted}$ and confidentiality as sensitivity levels ordered from public to critical.
When values are combined, their labels merge towards the conservative element, finding the greatest lower bound (meet) for integrity and the least upper bound (join) for confidentiality.
Concretely, combining a trusted with an untrusted value yields an untrusted one, while combining public with sensitive data carries the sensitive label.
The initial labels are assigned by the policy, which must be specified \emph{a priori}.
The policy declares locations of the mediated state and specific parties as trusted sources.
It assigns sensitivity levels to the locations and parties whose data is sensitive, and to each party a clearance.
Whether the principal's task is a trusted source is likewise a declaration of the policy.
All other content is considered untrusted and public.
\para{Provenance}
Determining whether an effect is authorized thus depends on the \emph{provenance} of its values.
The monitor, however, sees values only where they cross the boundary, i.e., the content the agent reads and the effects it emits, but not how one turns into the other.
Labels therefore enter with incoming events which carry the label of their source.
To build provenance, the monitor keeps a record of what the agent has read.
$T_i$ collects the values the agent has received from trusted sources up to event $e_i$, starting from the values the policy itself declares, and $S_i$ collects the sensitive values it has received.
For an incoming event $e_i$ from a trusted source, $T_i = T_{i-1} \cup \{v_i\}$, where $v_i$ is its content, and $S_i$ grows likewise.
$T_i$ is thus the set of values the agent could have obtained legitimately.
Hence, a value in an effect that stems from $T_i$ is justified by a trusted source, and any other value is not.
Both sets grow only with incoming events, and nothing the agent emits can add to $T_i$, as the agent's own output is exactly what the monitor cannot trust.
Since values pass through the agent's model and private state, which the monitor cannot observe, labels have to be re-established at the boundary.
A value $x$ in an effect is trusted only if it is \emph{derivable} from $T_i$: $D(x, T_i)$.
Sensitivity is checked in the opposite direction, since a missed match has opposite consequences on the two axes, as a value that is not recognized as trusted is deferred, whereas sensitive data that is not recognized would be released.
An effect therefore carries sensitive data if a value of $S_i$ occurs anywhere in it, such as a private key embedded in a URL.
We require $D$ to be sound against the injected agent of \cref{sec:threat}, which does not deliberately transform or otherwise mask values.
Tracking individual values through the inference of a language model remains an open problem, and approaches that avoid this assumption propagate labels coarsely through the agent's context instead~\cite{camel,fides}, which requires modifying the agent and taints much of what it emits (\cref{sec:discussion}).
For the same reason, we also require that a location the agent writes inherits the merged labels of everything the agent read before the write, so that state cannot launder labels between effects or across sessions.
Within a speculative scope, labels admitted by held effects stay local to the scope and become permanent only when they are committed (\cref{subsec:mechanism}).
\para{Policy}
When the monitor has to decide on an effect, it first consults the policy before relying on the oracle.
The policy answers three questions in order.
\emph{First, does an explicit rule of the operator forbid the effect?}
If so, such as for deleting a repository's history, the effect is refused.
\emph{Second, do the values that steer the effect stem from a trusted source?}
Not every value of an effect needs to be trusted: a summary of an e-mail legitimately contains the e-mail's text.
Only the values that steer the effect, such as its destination, its recipient, or the command it runs, must do so.
We write $G(e)$ for these \emph{gated} values and require $D(x, T_{i-1})$ for every $x \in G(e_i)$.
\emph{Third, is the data the effect carries cleared for its target?}
With $\ell(e)$ the join of the sensitivity levels of the values of $S_{i-1}$ that occur in $e$, and $\mathit{clr}(\mathit{obj}(e))$ the clearance of its target, we require $\ell(e_i) \sqsubseteq \mathit{clr}(\mathit{obj}(e_i))$.
Locations of the mediated state carry the highest clearance, since data written there does not leave it and keeps its label.
If both checks hold, the policy \emph{vouches} for the effect, and otherwise it defers the effect to the oracle:
\begin{equation}
  \delta(\sigma_{i-1}, e_i) =
  \begin{cases}
    \refuse  & \text{if a rule of the policy forbids } e_i,\\
    \release & \text{if the policy vouches for } e_i,\\
    \defer   & \text{otherwise.}
  \end{cases}
\end{equation}
For example, a request to an address named in the task that carries only public data is released, while the same request to an address found in an e-mail is deferred.
For a blocking effect, deferral cannot hold the effect, so the oracle decides it before it is sent.
\para{Oracle}
Deferred effects are decided by the \emph{decision oracle}.
The decision oracle can be the operator, another policy engine, or another model instance as in our evaluation (\cref{sec:implementation}).
The oracle receives a $\mathit{view}$ the monitor curates, consisting of the principal's task and the policy, the deferred effects with the provenance of their values, the trace that led to them, and the untrusted content they build on as evidence, and it returns a decision for each deferred effect:
\begin{equation}
  O\big(\mathit{view}(\sigma)\big) : \mathit{def}(\sigma) \rightarrow \{\release, \refuse\},
\end{equation}
where $\mathit{def}(\sigma)$ are the effects $\delta$ deferred, each accompanied by a justification that serves as the notice to the agent and the record of the decision.
The oracle's authority is bounded by construction as every effect in $\mathit{obs}(\sigma)$ was either released by $\delta$ or deferred by $\delta$ and released by $O$, so the oracle can release only what the policy deferred and can never override a refusal.
It is therefore the only point at which an untrusted value can reach a released effect or sensitive data an uncleared destination.
Hence the oracle represents the point of endorsement and declassification~\cite{sabelfeld_sands2009, myers2006robust}.
An oracle that refuses every deferred effect reduces the monitor to the strict monitor of \cref{sec:threat}, and any other oracle can only add releases.
Since its view contains untrusted content, the oracle is itself a target of the attacker (see \Cref{sec:discussion} for a discussion on the impact of the bandwidth of this channel).

\subsection{Implementation}
\label{sec:implementation}\label{sec:instantiation}
\begin{figure}[t]
\centering
\includegraphics[width=\columnwidth]{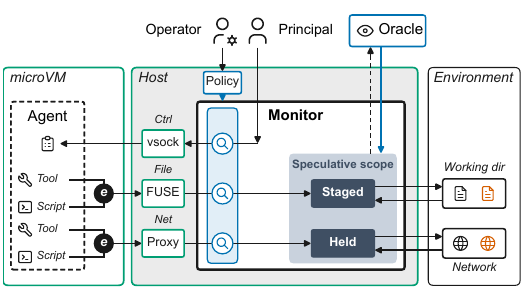}
\caption{Overview of \sysname{}.}
\label{fig:implementation}
\end{figure}

\para{Isolation and channels}
The agent operates within a Cloud Hypervisor~\cite{cloudhypervisor} microVM, which provides a hardware-enforced boundary without a shared kernel.
The working directory constitutes the mediated state and is exported to the guest via the $\mathsf{file}$ channel.
The guest's root filesystem, containing the agent's tool installations and credentials, remains outside the scope of mediation.
The $\mathsf{file}$ channel mounts the working directory as the read-only lower layer of an OverlayFS~\cite{overlayfs}, and all modifications go to a separate upper layer, so the working directory stays unchanged and the upper layer holds exactly the staged state.
The merged view of both layers is re-exported to the guest using a FUSE passthrough~\cite{fuse2017} and virtio-fs, ensuring that all file operations are executed by a user-space process on the host.
The $\mathsf{net}$ channel functions as a transparent TLS-terminating proxy implemented with mitmproxy~\cite{mitmproxy}.
A firewall rule redirects all outbound traffic to the proxy, and the proxy's certificate authority is embedded within the guest image.
Terminating TLS at the proxy separates mediation from the agent's TLS library, unlike eBPF hooks, which depend on specific TLS implementations~\cite{agentsight}.
File writes and unlinks are stageable, requests with state-changing methods stallable, and \texttt{GET} requests blocking, as we assume the agent requires their response.
Traffic directed to the model provider is decrypted, logged, and released, as the provider is considered an internal channel (\cref{subsec:confinement}).
The $\mathsf{ctrl}$ channel is the agent's terminal over vsock, through which the task arrives as the agent's start command and the response returns to the principal.
Since these channels are the only paths to the working directory, the network, and the principal, every effect crosses the monitor~(\ref{g:mediation}), and a tool call and a script the agent wrote reach it as the same event (\cref{fig:implementation}).
\para{Holding}
The upper layer of the overlay holds the writes of the open speculative scope, and a write-ahead log records a pre-image of each change, so that the scope can be rolled back.
Committing the scope merges its changes into the working directory, and aborting it discards them.
A held request is stalled at the proxy while the agent continues on a stub that replaces $\mathit{res}(e)$, and it is sent only when the scope commits.
The prototype keeps one speculative scope open at a time, as OverlayFS offers only a single writable upper layer per mount.
A further effect that would need holding therefore forces the open scope to resolve first, as does a state-changing request, content derived from a stub that is about to leave the boundary, or the end of the task.
When a held effect is refused, the agent receives an abort notice specifying the withheld effects, affirming that the remainder of the task continues, instructing the agent to proceed, and permitting re-issuance only for effects explicitly required by the task.
\para{Policy and provenance}
The policy is defined as a short file that declares the trusted and sensitive sources and the parties with their clearances, together with a small layer of rules.
Parties the policy does not name are deferred in all configurations.
Following \cref{subsec:policy}, provenance is re-established at the boundary by matching values.
We instantiate derivability as equality up to normalization,
\begin{equation}
  D(x, T_i) \iff \exists\, t \in T_i : \mathit{norm}(x) = \mathit{norm}(t),
\end{equation}
where $\mathit{norm}$ lowercases a value and removes its whitespace.
$T_0$ holds the values extracted from the principal's task, and each result from a declared trusted source adds its values.
For sensitivity, $S_i$ holds the tokens of the content read from sensitive sources, and an effect carries a sensitive value if one of these tokens occurs in it after normalization, which over-approximates towards deferral.
Matching misses values the agent transforms, and it cannot tell an attacker's value from an identical trusted one.
We discuss both limits in \cref{sec:discussion}.
\para{Decision oracles}
The oracle $O$ is invoked once per resolution of a speculative scope and, for a blocking request that the policy defers, before the request is sent.
Its view (\cref{subsec:policy}) presents the trace with untrusted content marked as data and adds the lines of earlier content that mention the values of the deferred effects.
The provenance of each value is stated as its origin, such as \emph{supplied by read\_file:bill.txt, not by the user's message}, rather than as a label, since a language model reads a label such as \emph{untrusted} as a verdict.
In our evaluation, this view is given to a separate language model instance without tools, queried at temperature zero, whose model is configured independently of the agent's.
Output that cannot be parsed and failed calls count as \refuse{}, so a failing oracle can only refuse.

\section{Evaluation}
\label{sec:evaluation}\label{sec:eval}
\providecommand{\miss}{\textcolor{red}{?}}
\begin{table*}[ht]
\providecommand{\grp}{\rule{0pt}{10.8pt}}%
\centering
\caption{Comparison of defenses on AgentDojo with three agent models (\%).}
\label{tab:agentdojo-main}
\small
\setlength{\tabcolsep}{4pt}
\begin{threeparttable}
\begin{tabular}{@{}ll rrr>{\columncolor{nrpbg}[\tabcolsep][\tabcolsep]}r rrr>{\columncolor{nrpbg}[\tabcolsep][\tabcolsep]}r rrr>{\columncolor{nrpbg}[\tabcolsep][0pt]}r@{}}
\toprule
& & \multicolumn{4}{c}{\textbf{GPT-4.1}} & \multicolumn{4}{c}{\textbf{Qwen3.6-35B}} & \multicolumn{4}{c}{\textbf{Llama-3.3-70B}} \\
\cmidrule(lr){3-6}\cmidrule(lr){7-10}\cmidrule(l){11-14}
\textbf{Level} & \textbf{Defense} & \textbf{ASR}$\downarrow$ & \textbf{U}$\uparrow$ & \textbf{U\textsubscript{b}}$\uparrow$ & \textbf{NRP}$\uparrow$ & \textbf{ASR}$\downarrow$ & \textbf{U}$\uparrow$ & \textbf{U\textsubscript{b}}$\uparrow$ & \textbf{NRP}$\uparrow$ & \textbf{ASR}$\downarrow$ & \textbf{U}$\uparrow$ & \textbf{U\textsubscript{b}}$\uparrow$ & \textbf{NRP}$\uparrow$ \\
\midrule
--        & No defense         & 47.0 & 59.8 & 82.7 & 43.8 & 28.3 & 72.0 & 87.0 & 62.4 & 32.3 & 24.6 & 29.8 & 20.2 \\
\grp Model     & PI detector~\cite{agentdojo} & 9.5 & 24.0 & 40.1 & 36.3 & 2.9 & 29.8 & 50.5 & 49.0 & 4.0 & 17.0 & 20.1 & 19.3 \\
\grp Prompt    & Prompt sandwiching~\cite{agentdojo} & 44.6 & 63.7 & \textbf{87.6} & 48.5 & 14.4 & \textbf{77.1} & \textbf{86.3} & 73.9 & 22.5 & 26.9 & 30.8 & 23.9 \\
          & Spotlighting~\cite{spotlighting} & 49.2 & 54.2 & 81.7 & 41.5 & 30.8 & 69.7 & 83.8 & 58.0 & 28.7 & 23.9 & \textbf{38.0} & 27.1 \\
\grp Framework & Progent~\cite{progent} & 11.5 & 57.6 & 78.2 & 69.2 & 5.4 & 70.2 & 82.1 & 77.7 & 8.0 & 32.0 & 31.8 & 29.3 \\
          & CaMeL~\cite{camel} & \textbf{0.2} & 40.7 & 38.0 & 37.9 & \textbf{0.4} & 37.0 & 40.4 & 40.2 & \textbf{0.0} & 26.0 & 26.5 & 26.5 \\
\grp System    & \sysname{}         & 2.7 & \textbf{69.5} & 85.1 & \textbf{82.8} & 1.3 & 76.5 & 84.6 & \textbf{83.5} & 2.4 & \textbf{34.0} & 32.3 & \textbf{31.5} \\
\bottomrule
\end{tabular}
\begin{tablenotes}[flushleft]\footnotesize
\item ASR: attack success rate; U: utility under attack; U\textsubscript{b}: benign utility; NRP: net resilient performance (\cref{subsec:eval-setup}).
\end{tablenotes}
\end{threeparttable}
\end{table*}

We compare \sysname{} against existing defenses across three benchmarks and three agent models, and analyse the contribution of its components and evaluate the overhead \sysname{} introduces.
The evaluation is structured along four questions.
\Cref{subsec:eval-effectiveness} asks whether \sysname{} stops injection attacks while preserving utility, compared to prompt-, model-, and framework-level defenses.
\Cref{subsec:eval-decisions} examines how much the policy decides on its own and what reaches the oracle.
\Cref{subsec:eval-ablations} isolates what deferral and the oracle contribute, and which model choices drive security and utility.
\Cref{subsec:eval-cost} quantifies what \sysname{} adds in tokens, model calls, and latency.
\subsection{Setup}
\label{subsec:eval-setup}
\para{Benchmarks}
We evaluate on AgentDojo~\cite{agentdojo}, the standard benchmark for indirect prompt injection, and on two extensions: AgentDyn~\cite{agentdyn}, whose legitimate actions are not known in advance, and AutoDojo~\cite{autodojo}, whose attacker adapts its injections to the defense.
\par
\emph{AgentDojo} simulates four application suites with 97 user tasks and 629 injection cases.
Each injection places the attacker's goal into content the agent reads while completing a user task.
We use its \texttt{important\_instructions} attack, against which CaMeL~\cite{camel} and Progent~\cite{progent} also report their results.
\emph{AgentDyn} contains 60 open-ended tasks across three domains with a total of 560 injection cases.
Here the agent has to continue planning as it uncovers information, since the set of legitimate actions is not known from the initial task alone.
Lastly, \emph{AutoDojo} extends three of the four AgentDojo suites (389 injection cases) with an adaptive attacker that uses an LLM to rewrite each injection over six rounds to bypass the defense.
This attacker is weaker than the one we assume in \cref{sec:threat} as it is black-box and can only observe whether an attack succeeds. So its results are a lower bound on adaptive attack success.
\para{Baselines}
We compare \sysname{} against defenses at three levels: the prompt, the model, and the agent framework.
\par
\emph{Prompt-level} defenses only change what the agent reads.
Spotlighting~\cite{spotlighting} marks untrusted content as data, and prompt sandwiching~\cite{schulhoff2023sandwich} repeats the user's task after it.
\emph{Model-level} defenses use a separate model to detect injected instructions in tool outputs before they reach the agent.
We compare against the PI detector~\cite{protectai2024deberta}, a fine-tuned classifier that all three benchmarks report.
\emph{Framework-level} defenses restrict which actions the agent can take.
Progent~\cite{progent} enforces a tool-call policy that a language model derives from the task, and CaMeL~\cite{camel} executes a program generated from the task under information-flow policies.
We run CaMeL with its security policies enforced and use GPT-4.1 as the auxiliary model for Progent, the same model as \sysname{}'s oracle.
Lastly, on AgentDojo we run all defenses in our own harness, while on AgentDyn and AutoDojo we report the values published with each benchmark.
\para{Models and harness}
Unless otherwise specified, the agent is GPT-4.1\footnote{gpt-4.1-2025-04-14} at temperature zero with a concise, task-neutral system prompt.
On AgentDyn and AutoDojo, we use the agent models of their original evaluations, GPT-4o and GPT-4o-mini respectively, so that our results are comparable to originally reported values.
On AgentDojo, we additionally evaluate Qwen3.6-35B\footnote{\href{https://huggingface.co/Qwen/Qwen3.6-35B-A3B/tree/995ad96eacd98c81ed38be0c5b274b04031597b0}{Qwen/Qwen3.6-35B-A3B}} and Llama-3.3-70B\footnote{\href{https://huggingface.co/meta-llama/Llama-3.3-70B-Instruct/tree/6f6073b423013f6a7d4d9f39144961bfbfbc386b}{meta-llama/Llama-3.3-70B-Instruct}} as agents.
Each benchmark defines the task through the user prompt and scores its completion.
All defenses run without human intervention, so a blocked action never receives a user's approval as a fallback and each configuration is run once.
\para{\sysname{}
integration}
As the benchmarks simulate tools as in-memory objects, there is no isolation boundary to mediate for \sysname{}.
We therefore replace the benchmark's tool executor with an adapter that turns each simulated tool call into an outgoing event and its result into an incoming one, and add one hook after the model's turn that resolves an open scope when the agent finishes.
Policy, provenance tracking, the speculative state machine, and the oracle run on these events unchanged.
Only the messages that carry a stub or an abort notice back to the agent are produced by the adapter.
However, the events differ, for a simulated tool call names its tool and arguments, whereas the deployed boundary observes file operations and requests.
The policy therefore declares, for each tool, whether it is reversible and which argument names its destination.
The policies declare no refusing rule, as such a rule encodes knowledge of an attack and blocks it by construction, so the attack success we report would reflect our foresight rather than the monitor.
In a deployment, such rules are what bounds the oracle, as no decision of the oracle can release an effect a rule refuses.
Unless otherwise specified, the oracle is GPT-4.1 and decides each deferred effect against the user's task.
\para{Metrics}
We evaluate each defense by its \emph{attack success rate} (ASR), defined as the percentage of injection cases in which the agent completes the attacker's goal.
Compared against the undefended agent's ASR, this shows how many attacks a defense prevents.
To capture the cost to the agent's usefulness, we measure its \emph{benign utility}~$U_b$, the percentage of user tasks completed when no attack is present, and its \emph{utility under attack}~$U$, the percentage of user tasks completed in the injection cases.
To quantify the trade-off between security and utility, we report the \emph{net resilient performance} (NRP) proposed by Zhang \emph{et al.}~\cite{NRP:DBLP:conf/iclr/ZhangHMYWZWZ25} as the main metric:
\begin{equation*}
    \mathrm{NRP} = U_{b} \times (1-\mathrm{ASR}).
\end{equation*}
\subsection{Effectiveness}
\label{subsec:eval-effectiveness}
\para{AgentDojo}
\label{subsec:eval-agentdojo}
With GPT-4.1 as the agent, \sysname{} reaches the highest NRP of all defenses, 82.8 against 69.2 for Progent, the second best defense (\cref{tab:agentdojo-main}).
It reduces attack success from 47.0\% to 2.7\%, with no successful attack in the Banking and Workspace suites, while its benign utility of 85.1\% matches that of the undefended agent (82.7\%).
Utility under attack rises from 59.8\% to 69.5\%, the highest of all defenses.
The other defenses trade one for the other: prompt-level defenses preserve utility but barely reduce attack success, whereas the model-level detector and CaMeL reduce attack success but complete fewer than half of the benign tasks.
The picture holds for the two open-weight agents, with the oracle still GPT-4.1: \sysname{} reduces attack success from 28.3\% to 1.3\% for Qwen3.6-35B and from 32.3\% to 2.4\% for Llama-3.3-70B.
With both agents, \sysname{} again reaches the highest NRP (83.5 and 31.5), with benign utility close to that of the undefended agent (84.6\% against 87.0\% for Qwen3.6-35B, 32.3\% against 29.8\% for Llama-3.3-70B). Notably, Llama completes few tasks even without a defense, so its utility is low throughout.
An undefended agent that follows an injection often abandons the user's task and when \sysname{} refuses the injected effect, it notifies the agent that the effect was not executed, that the remainder of the user's request remains valid, and instructs the agent to proceed. This recovers some utility.
\para{AgentDyn}
\label{subsec:eval-agentdyn}
On open-ended tasks (\cref{tab:agentdyn}), \sysname{} reduces attack success to 1.7\%, against 37.8\% reported for the undefended agent, and achieves the highest NRP (50.8 against 43.6 for prompt sandwiching).
The other defenses diverge further than on AgentDojo.
Prompt-level defenses keep their utility but reduce the ASR by at most 10.2 points compared to the undefended agent.
The PI detector and Progent reach similarly low attack success, but complete almost no tasks under attack (0.6\% and 5.8\%).
CaMeL is excluded here as it fixes its program before the task begins.
\sysname{} keeps its benign utility close to that of the undefended agent (51.7\% against 53.3\%), but open-ended tasks come at a cost in utility under attack (37.1\% against 55.5\%).
This cost stems from held effects that later steps depend on, and it grows with each hold a task has to continue past.
Tasks that continue past a single held effect retain 90\% of their utility, while those that continue past two or more retain 62\%. This is discussed further in \cref{sec:discussion}.
\para{AutoDojo}
\label{subsec:eval-adaptive}
\sysname{} remains effective against an adaptive attacker where injections optimized against it succeed in 8.8\% of the cases, compared to 52.4\% without a defense (\cref{tab:autodojo}).
Together with Progent, it retains the highest NRP, as both keep attack success in single digits without giving up clean utility.
Prompt-level defenses leave between 36\% and 47\% of the attacks successful, and the PI detector reduces attack success only to 15.4\% while completing fewer than half of the clean tasks.
The attacks that still succeed target tasks that delegate authority to a document, which the oracle reads as authorized by the user.
\subsection{Policy and Oracle Decisions}
\label{subsec:eval-decisions}
\providecommand{\grp}{\rule{0pt}{10.8pt}}%
\begin{table}[t]
\centering
\caption{Comparison of defenses on AgentDyn with GPT-4o as the agent (\%).}
\label{tab:agentdyn}
\small
\setlength{\tabcolsep}{4.5pt}
\begin{threeparttable}
\begin{tabular}{@{}llrrr>{\columncolor{nrpbg}[\tabcolsep][0pt]}r@{}}
\toprule
\textbf{Level} & \textbf{Defense} & \textbf{ASR}$\downarrow$ & \textbf{U}$\uparrow$ & \textbf{U\textsubscript{b}}$\uparrow$ & \textbf{NRP}$\uparrow$ \\
\midrule
--        & No defense$^{\dagger}$          & 37.8 & 55.5 & 53.3 & 33.2 \\
\grp Model     & PI detector$^{\dagger}$         &  \textbf{0.9} &  0.6 &  0.0 &  0.0 \\
\grp Prompt    & Prompt sandwiching$^{\dagger}$  & 31.2 & \textbf{56.1} & \textbf{63.3} & 43.6 \\
          & Spotlighting~\cite{spotlighting}$^{\dagger}$        & 27.6 & 52.2 & 55.0 & 39.8 \\
\grp Framework & Progent~\cite{progent}$^{\dagger}$             &  1.7 &  5.8 &  6.7 &  6.6 \\
\grp System    & \sysname{}                      &  1.7 & 37.1 & 51.7 & \textbf{50.8} \\
\bottomrule
\end{tabular}
\begin{tablenotes}[flushleft]\footnotesize
\item $^{\dagger}$\,Reported by AgentDyn.
\end{tablenotes}
\end{threeparttable}
\end{table}

\begin{figure}[t]
\centering
\includegraphics[width=\columnwidth]{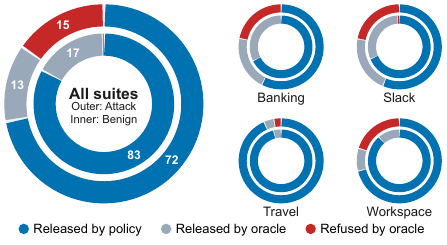}
\caption{Decisions per effect on AgentDojo (GPT-4.1, \%).}
\label{fig:decisions}
\end{figure}

The design bounds the oracle to the effects the policy defers (\cref{subsec:policy}), so how much of the monitor's behavior rests on the oracle depends on how often the policy defers.
As the policy decides without a model call, this share also determines the token and latency overhead of \sysname{} (\cref{subsec:eval-cost}).
We therefore examine which component decides the effects of a run, and how the oracle decides the effects that reach it.
\Cref{fig:decisions} shows where the effects of the AgentDojo evaluation (\cref{tab:agentdojo-main}) with GPT-4.1 are decided.
The policy directly releases most of the effects, 83\% of the benign effects and 72\% under attack, so only a minority ever reaches the oracle.
In the benign runs, the oracle releases all but one effect it decides, the single refusal is the only false positive.
Under attack, the policy defers more effects, and the oracle refuses more than half of them.
Most deferrals stem from values in an effect whose provenance the policy cannot trace to a trusted source.
\subsection{Oracle and Model Choice}
\label{subsec:eval-ablations}
\para{Strict monitor}
\label{subsec:eval-ablation}
To measure what deferral and the oracle contribute, we disable both, so that every effect the policy defers is refused.
This is the strict monitor that an oracle refusing everything would yield (\cref{subsec:policy}).
Its attack success stays within the variation of a single run, so trusting the oracle costs no measurable security on AgentDojo.
Its benign utility, however, halves from 85.1\% to 42.7\%, and its utility under attack drops from 69.5\% to 34.7\%.
The policy thus provides the security, while deferral and the oracle together recover the utility that a strict monitor gives up.
\para{Agent and oracle models}
\label{subsec:eval-models}
To distinguish the roles of the two models, the agent and the oracle are varied independently across three sizes within the Qwen3.5 family on the AgentDojo Banking suite (\cref{tab:model-grid}). When the oracle is fixed, the 9B agent consistently achieves the lowest NRP across all oracles (68.8 to 81.2). However, all pairings remain 18.0 to 52.4 points above the undefended agent. When the agent is fixed, a 9B oracle reduces NRP by 10.7 to 12.5 points for the 27B and 9B agents, but not for the 35B agent. The 35B and 27B oracles perform similarly, differing by no more than 1.7 points. When the agent and oracle are the same model, NRP ranges from 93.1 for the 27B model to 68.8 for the smallest. Therefore, deployment does not require the strongest model as the oracle. A mid-sized oracle is sufficient, while a weaker agent consistently lowers NRP across all oracles.
\subsection{Cost}
\label{subsec:eval-cost}
\begin{table}[t]
\providecommand{\grp}{\rule{0pt}{10.8pt}}%
\centering
\caption{Comparison of defenses on AutoDojo with GPT-4o-mini as the agent (\%).}
\label{tab:autodojo}
\small
\setlength{\tabcolsep}{4.5pt}
\begin{threeparttable}
\begin{tabular}{@{}llrr>{\columncolor{nrpbg}[\tabcolsep][0pt]}r@{}}
\toprule
\textbf{Level} & \textbf{Defense} & \textbf{ASR}$\downarrow$ & \textbf{U\textsubscript{b}}$\uparrow$ & \textbf{NRP}$\uparrow$ \\
\midrule
--        & No defense$^{\dagger}$ & 52.4 & 66.7 & 31.7 \\
\grp Model     & PI detector$^{\dagger}$ & 15.4 & 47.4 & 40.1 \\
\grp Prompt    & Prompt sandwiching$^{\dagger}$ & 36.3 & 59.6 & 38.0 \\
          & Spotlighting~\cite{spotlighting}$^{\dagger}$ & 46.8 & 59.6 & 31.7 \\
\grp Framework & Progent~\cite{progent}$^{\dagger}$ & \textbf{7.7} & 63.2 & 58.3 \\
\grp System    & \sysname{} & 8.8 & \textbf{65.3} & \textbf{59.6} \\
\bottomrule
\end{tabular}
\begin{tablenotes}[flushleft]\footnotesize
\item $^{\dagger}$\,Reported by AutoDojo.
\end{tablenotes}
\end{threeparttable}
\end{table}

Because the policy decides without a model call, the token overhead, which is 17.7\% on benign runs and 33.6\% under attack, corresponds to the number of deferrals and is fully attributable to oracle calls.
For comparison, CaMeL reports approximately 2.8 times the input and 2.7 times the output tokens of a single-mode agent~\cite{camel}, while AttriGuard incurs about twice the tokens and three times the latency~\cite{attriguard}.
\par
The isolation mechanism introduces only a manageable overhead.
Across 200 cycles of create, write, fsync, read, and unlink operations on 4 KB files, repeated three times, the filesystem channel adds an average of $9.62$ ms per cycle or approximately $1.9$ ms per operation.
For 50 allowed requests, the proxy adds an average of 1.15 ms per request.
Compared to the agent's model inference latency per step~\cite{openrouter}, this overhead is negligible.

\section{Discussion and Limitations}
\label{sec:discussion}
\sysname{} guarantees confinement for any agent and any oracle, whereas authorization depends on both the policy and the oracle and can only be measured (\cref{sec:threat}).
The evaluation attributes the measured security to the policy (\cref{subsec:eval-ablation}), which is expected, because an injection that introduces its own value, such as a recipient or a URL, fails the integrity check unless its value matches a trusted one.
The oracle determines only what the policy has deferred.
The limitations of both mechanisms are discussed below.
\para{Semantic gap}
At the isolation boundary, the monitor observes file operations and requests instead of tool names and arguments.
While this placement provides complete mediation (\cref{sec:threat}), it introduces a semantic gap.
A policy defined over tools can associate a rule with an operation and the role of each argument~\cite{progent,agentspec}, whereas a policy defined over events refers only to a path or destination and the values carried by an effect, and can thus express only a subset of these rules.
Any rule that the policy cannot express must be deferred, so the oracle has to recover it from the task and the ordered trace.
Existing benchmarks expose every action as a named tool call, to our adapter as well (\cref{subsec:eval-setup}), and none composes an attack from individually legitimate steps, so measuring this class of defense requires a benchmark in which actions correspond to actual system effects.
\para{Provenance precision}

Any defense that permits untrusted content to reach a model must contend with the inability to precisely track data flow through inference.
\sysname{} considers a broader black box approach, encompassing the model, its framework, and associated tools, and therefore shares this limitation with existing defenses.
Provenance is re-established by matching values at the isolation boundary (\cref{sec:implementation}).
If an attacker's value is byte-identical to a trusted value, it is assigned the same provenance.
In contrast, if the agent transforms a value, it is no longer recognized, which is why the threat model excludes deliberate transformation (\cref{sec:threat}).
Similarly, sensitivity declared for an object, such as a file path, does not persist if the agent moves the data to another object before reading it.
Propagating the labels of all data read by the agent to every subsequent effect would maintain soundness but would defer nearly all effects after the first untrusted read.
Label propagation within the agent~\cite{fides} provides greater precision by tracking each value through the framework into and out of the model's context at the cost of modifying the agent.
\para{Decision oracle vulnerability}
The attacker can influence the oracle solely through its view, which must present the untrusted content underlying a deferred effect as evidence.
While this design constrains the attack channel as each span carries its provenance, the content under review may still attempt to persuade the oracle, particularly when the task delegates authority to a document and the oracle interprets the document's instructions as those of the principal, which enables the remaining adaptive attacks to succeed (\cref{subsec:eval-adaptive}).
An attacker familiar with the oracle and its view operates through the same channel, as it controls neither the task nor the policy and can, at most, cause a deferred effect to be released.
Nevertheless, such an attacker can optimize content against the oracle more effectively than a black-box attacker~\cite{nasr2025attacker}, so these results do not bound the attacker's success rate.
This limitation is inherent to any approach that interprets untrusted content to decide, and is independent of the present design.
Reducing the capacity of the channel between the attacker and the oracle by, for example, only allowing untrusted content as restricted typed data~\cite{jacob2025typed} is likely to reduce but not mitigate the attack vector entirely.
\para{Repeated deferral}
\begin{table}[t]
\providecommand{\miss}{\textcolor{red}{?}}
\centering
\caption{NRP with agent and oracle models varied independently (\%, AgentDojo Banking).}
\label{tab:model-grid}
\small
\setlength{\tabcolsep}{5pt}
\begin{threeparttable}
\begin{tabular}{@{}lrrrr@{}}
\toprule
 & & \multicolumn{3}{c}{\textbf{Oracle}} \\
\cmidrule(l){3-5}
\textbf{Agent} & \textbf{None} & \textbf{35B} & \textbf{27B} & \textbf{9B} \\
\midrule
Qwen3.5-35B & 62.6 & 86.3 & 87.5 & 92.5 \\
Qwen3.5-27B & 40.7 & 93.1 & 93.1 & 80.6 \\
Qwen3.5-9B  & 50.8 & 79.5 & 81.2 & 68.8 \\
\bottomrule
\end{tabular}
\begin{tablenotes}[flushleft]\footnotesize
\item Qwen3.5-35B-A3B (mixture-of-experts). None: the undefended agent.
\end{tablenotes}
\end{threeparttable}
\end{table}

Deferral is essential for preserving utility and, consequently, the monitor's usability, especially during extended sessions where repeated application leads to cumulative costs.
When the oracle releases an effect, it declassifies only that specific effect and retains the original labels of its associated values.
As a result, any subsequent effect involving the same untrusted value is subject to repeated deferral.
As untrusted content increases, deferrals occur more frequently, each withholding an effect that subsequent steps may require, and each decision is made based on an increasingly long trace.
This process contributes to utility loss in open-ended tasks (\cref{subsec:eval-agentdyn}).
Partitioning the trace according to provenance would restrict a deferral to the effects within a single lineage, providing the oracle with a shorter and more pertinent trace.
In this scenario, a release could apply to subsequent effects of the same lineage directed toward the same destination, albeit at the expense of a less granular declassification.
Both approaches require a formal definition of when two values share a lineage, which string matching alone cannot provide.
\para{Applicability}
While we evaluate \sysname{} on a relatively narrow scope of injection cases following the established benchmarks, the conceptual applicability of this architecture is much broader.
Firstly, because every effect crosses the monitor before it is released, the boundary also yields a complete record of the agent's effects and of each decision with its justification.
Additionally the oracle provides a point at which a human can intervene before an effect becomes irreversible.
This directly supports the record-keeping and human-oversight obligations that the EU AI Act places on high-risk systems (Art.~12 and~14)~\cite{euaiact}, although compliance depends on the deployment as a whole.
\par
Beyond observation, the guarantees of confinement do not depend on why the agent emits a malicious effect.
They hold equally when an effect stems from an injection, a misaligned or backdoored model, a poisoned tool, or a plain mistake, as the threat model already treats the agent as potentially fully adversarial (\cref{subsec:confinement}).
This matters beyond injection attacks, as recent incidents show agents compromising infrastructure far outside their task~\cite{hf_incident,metr_incident}.
Authorization generalizes partially, as the policy judges where the values of an effect come from and not who introduced them, so an effect steered by a value that no trusted source provides is deferred whether an injection or the agent itself produced it.
However, scenarios that consider, for example, a misaligned model need to account for values the model deliberately transforms to hide their origin, which our authorization assumptions exclude due to the limits of provenance tracking discussed above (\cref{sec:threat}).

\section{Related Work}
\label{sec:related}
We categorize prior work based on the enforcement level and the capabilities enabled at each point.
\para{Prompt- and model-level defenses}
Prompt- and model-level defenses either mark or separate untrusted content at inference time \cite{spotlighting}, or encode an ordering of instruction sources into the model weights \cite{instructionhierarchy}.
Guardrail middleware filters both incoming and outgoing content \cite{nemoguardrails}, while detectors classify injected text before it reaches the agent \cite{llamafirewall}.
These defenses operate solely on text and do not observe the agent's impact on the system.
On AgentDojo, they either maintain the attack rate near the undefended baseline or disable the agent entirely.
LlamaFirewall's AlignmentCheck \cite{llamafirewall} audits the agent's trace against the user's goal and is most similar to the oracle described here.
However, it evaluates without provenance and lacks the ability to hold or reverse actions.
\para{Framework-level defenses}
Structural defenses primarily enforce policies at the agent's tool dispatch layer.
CaMeL, for example, separates a privileged planner from a quarantined model to prevent untrusted data from influencing control flow \cite{camel}.
Another approach propagates information-flow labels through the planner or through inter-agent communication \cite{fides, safeflow}.
Propagating labels through the planner's context is sound within the label algebra and can detect decision-influence attacks that are outside the present threat model, but this method often results in significant over-tainting.
Once the planner accesses untrusted content, all subsequent actions inherit its label, leading to denial of actions even when they are part of the intended task, such as appending a fetched paper's summary to a local file.
Privilege-based defenses restrict the calls an agent may make, with policies over tool names and arguments written in a domain-specific language \cite{agentspec}, derived from the task by an LLM and refined during execution, with any expansion left to user discretion \cite{progent}.
Alternatively, policies may be expressed as Datalog rules over a causal dependency graph of the agent's actions \cite{forge} and generalized to probabilistic predicates with a sound bound on violation probability \cite{probverif}.
Attribution-based defenses verify whether a call was proposed due to injected data.
AgentArmor constructs a program-dependence graph from the trace and performs type checking \cite{agentarmor}, while AttriGuard re-executes the agent's model on a modified history in which an LLM rewrites untrusted observations from commands into third-person reports, blocking the call if the replay no longer proposes it~\cite{attriguard}.
These mechanisms provide more detailed per-call semantics than \sysname{}, but they rely on agent cooperation, whether through a replaced planner, an instrumented framework, or control over the inference loop.
As a result, their coverage is limited to the declared tool interface.
FORGE, for example, excludes raw sockets, standard I/O, and shell effects from its mediation, and its probabilistic generalization deems agents' ad-hoc scripts impossible to model in advance \cite{probverif}.
Additionally, most of these defenses permit or deny actions at the time of the call, which means that actions influenced by untrusted content are refused even when they are required by the task.
Progent's approval step is an exception, but in the absence of human intervention, it also results in denial.
Enforcement at the tool interface and \sysname{} are therefore complementary, since the former provides rich semantics with interface-relative coverage, while the latter ensures complete coverage at the isolation boundary, with semantics reconstructed at resolution.
\para{Transactions and rollback}
Transactional runtimes aggregate the tool calls of a task into a unit, maintaining reversible mutations in a shadow state while external effects are staged until the unit is validated \cite{cordon}.
Each effect is settled based on whether it can be buffered, compensated, or cannot be undone \cite{atomix}, or workflows are structured as stages with compensating steps \cite{sagallm}.
Recovery-oriented systems, by contrast, reverse effects after they occur, using snapshots taken before commands of uncertain effect \cite{yan_sandbox}, per-tool compensation actions \cite{goex}, or checkpoints and branches of the framework's state \cite{agentgit}.
The primary distinction between these approaches and \sysname{} lies in the transaction boundary and the commit decision process.
Runtimes and recovery systems interpose at tool dispatch or within the framework, so effects produced outside the declared interface are not included in the transaction.
Generally, commits are based on progress, failure, or operator decision, rather than on the provenance of values.
\sysname{} instead places the transaction boundary at the isolation boundary and lets provenance and the oracle's decision on the ordered trace determine the commit.
\para{Isolation and system-level monitoring}
Recent research suggests that agents should be secured similarly to other systems, treating the model as an untrusted component and enforcing invariants at the system level~\cite{systemsproblem}.
Observability tools already monitor agents at this level.
AgentSight, for example, traces system calls and encrypted traffic using eBPF, enabling an observer model to correlate the agent's LLM traffic with its system activity~\cite{agentsight}.
This tool detects rather than enforces, and its visibility into encrypted traffic depends on hooks into a specific TLS library.
General runtime security monitors and agent telemetry platforms record events without agent semantics or enforcement~\cite{falco,langsmith}.
Isolation is provided by user-space kernels, microVMs, and hosted agent sandboxes~\cite{gvisor,firecracker2020,e2b,daytona,llm_sandbox}.
These mechanisms protect the host from the agent, but the attacks under consideration remain within the scope authorized by the sandbox, as an injected agent misuses only the files and network access granted to it.
\sysname{} leverages the same isolation mechanisms and introduces additional mediation.
\para{Information-flow}
The labeling approach in this work follows Denning's lattice model \cite{denning1976}.
Operating system-level information-flow control enforces labels on processes managed by the kernel \cite{histar2006,flume2007}, and dynamic taint tracking monitors data through code that does not cooperate \cite{newsome_song2005,taintdroid2010}.
TaintDroid is the most closely related system, but it instruments the runtime it tracks \cite{taintdroid2010}.
In contrast, \sysname{} observes only the input and output of a component whose internals it cannot inspect, resulting in labels that are approximate by design.

\section{Conclusion}
\label{sec:conclusion}
Providing adequate isolation and authorization mechanisms for LLM agents becomes increasingly important as the capabilities and autonomy of widely deployed agents continue to grow.
With \sysname{}, we combine system-level isolation with a monitor that authorizes each effect of the agent at the isolation boundary.
Since every effect of an unmodified agent crosses this interface, whether it stems from a tool call or from a script the agent wrote, complete mediation becomes a property of the isolation layer rather than of the agent's cooperation.
Authorization rests on a declarative policy, which releases an effect when the values that steer it stem from a trusted source, such as the principal's task.
Additionally, the monitor can hold effects on both the file and the network channel, so that it can defer the decision on an effect the policy cannot vouch for while the agent continues speculatively.
Before such an effect becomes irreversible, a decision oracle, instantiated in our evaluation as a language model, decides whether to release or refuse it based on the agent's trace.
We show that \sysname{} achieves the highest net resilient performance among evaluated defenses across three benchmarks and three agent models.
This results from combining the security provided by the policy with the utility recovered by letting the oracle decide deferred effects later, when more of the trace is available.
These results highlight that the trade-off between security and utility commonly reported for agent defenses can be improved through deferral and speculative execution.
Overall, while we evaluate \sysname{} on common prompt injection benchmarks, confinement at the boundary holds regardless of why an agent emits an effect, and we see such a monitor as a foundation for overseeing increasingly autonomous agents beyond prompt injection.

\section{Open Science}
The source code of \sysname{}, including all components, is available at \url{https://github.com/UzL-ITS/Kairos}. The commercial coding agent utilized in the deployed stack is closed-source and therefore not included.
Because the evaluation relies on hosted models that providers may alter or discontinue, the released logs enable verification of the reported results without the need for re-execution.

\section{LLM Usage Considerations}
Large language models are central to this study. They function as agents under test, as the decision oracle, and, in the adaptive study, as the attacker (\cref{sec:eval}). Their use is therefore inherent to the research question, and we did not train models or collect data from human subjects.
The primary experiments employ open-weight and commercial models, accessed via OpenRouter under fixed model identifiers.
Because providers may update the models behind these identifiers, results may not be exactly reproducible, so we release all logs with our artifact.
To limit the volume of queries, we reuse results reported by prior work where the setup matches, as marked in \cref{sec:eval}.
During the development of this paper, we used LLMs for coding assistance.
LLMs were used for editorial purposes in this manuscript, and all outputs were inspected by the authors to ensure accuracy and originality.
The ideas and results presented in this work are solely those of the authors.

\section*{Acknowledgments}
This work was funded by the German Federal Ministry of Research, Technology and Space (BMFTR) through the projects AnoMed-II and AnoMed-KI.
Lorenzo Cavallaro was partially supported by a Google Academics Research Award.

\bibliographystyle{IEEEtran}
\bibliography{references}

\appendices
\section{Additional Details}
\begin{table*}[t]
\providecommand{\grp}{\rule{0pt}{10.8pt}}%
\centering
\caption{AgentDojo per suite for the three agents (\%). Columns as in \cref{tab:agentdojo-main}; means are unweighted over the four suites.}
\label{tab:main}
\small
\setlength{\tabcolsep}{4pt}
\begin{threeparttable}
\begin{tabular}{@{}ll rr rr rr rr rrr>{\columncolor{nrpbg}[\tabcolsep][0pt]}r@{}}
\toprule
& & \multicolumn{2}{c}{\textbf{Banking}} & \multicolumn{2}{c}{\textbf{Slack}} & \multicolumn{2}{c}{\textbf{Travel}} & \multicolumn{2}{c}{\textbf{Workspace}} & \multicolumn{4}{c}{\textbf{Mean}} \\
\cmidrule(lr){3-4}\cmidrule(lr){5-6}\cmidrule(lr){7-8}\cmidrule(lr){9-10}\cmidrule(l){11-14}
\textbf{Level} & \textbf{Defense} & ASR & U & ASR & U & ASR & U & ASR & U & ASR & U & U\textsubscript{b} & NRP \\
\midrule
\multicolumn{14}{@{}l}{\textit{GPT-4.1}} \\
--        & No defense          & 43.1 & 83.3 & 88.6 & 60.0 & 21.4 & 67.9 & 35.0 & 27.9 & 47.0 & 59.8 & 82.7 & 43.8 \\
\grp Model     & PI detector         & 3.5 & 30.6 & 16.2 & 30.5 & \textbf{0.0} & 13.6 & 18.3 & 21.2 & 9.5 & 24.0 & 40.1 & 36.3 \\
\grp Prompt    & Prompt sandwiching  & 36.8 & \textbf{86.1} & 74.3 & \textbf{62.9} & 37.1 & 62.1 & 30.0 & 43.8 & 44.6 & 63.7 & \textbf{87.6} & 48.5 \\
          & Spotlighting        & 45.1 & 79.2 & 81.9 & 54.3 & 28.6 & 58.6 & 41.2 & 24.6 & 49.2 & 54.2 & 81.7 & 41.5 \\
\grp Framework & Progent             & 9.7 & 72.2 & 7.6 & 44.8 & 17.9 & 71.4 & 10.8 & 42.1 & 11.5 & 57.6 & 78.2 & 69.2 \\
          & CaMeL               & \textbf{0.0} & 36.1 & \textbf{1.0} & 39.0 & \textbf{0.0} & 33.6 & \textbf{0.0} & 54.2 & \textbf{0.2} & 40.7 & 38.0 & 37.9 \\
\grp System    & \sysname{}          & \textbf{0.0} & 82.6 & 2.9 & 61.0 & 7.9 & \textbf{74.3} & \textbf{0.0} & \textbf{60.0} & 2.7 & \textbf{69.5} & 85.1 & \textbf{82.8} \\
\midrule
\multicolumn{14}{@{}l}{\textit{Qwen3.6-35B}} \\
--        & No defense          & 27.8 & 77.8 & 64.8 & 69.5 & 15.0 & 64.3 & 5.8 & 76.2 & 28.3 & 72.0 & 87.0 & 62.4 \\
\grp Model     & PI detector         & \textbf{0.0} & 36.8 & 6.7 & 33.3 & \textbf{0.0} & 3.6 & 5.0 & 45.4 & 2.9 & 29.8 & 50.5 & 49.0 \\
\grp Prompt    & Prompt sandwiching  & 17.4 & 76.4 & 36.2 & 66.7 & 2.1 & \textbf{87.1} & 2.1 & 78.3 & 14.4 & \textbf{77.1} & \textbf{86.3} & 73.9 \\
          & Spotlighting        & 35.4 & 77.8 & 58.1 & \textbf{70.5} & 20.0 & 58.6 & 9.6 & 72.1 & 30.8 & 69.7 & 83.8 & 58.0 \\
\grp Framework & Progent             & 8.3 & 74.3 & 5.7 & 61.9 & 4.3 & 80.0 & 3.3 & 64.6 & 5.4 & 70.2 & 82.1 & 77.7 \\
          & CaMeL               & \textbf{0.0} & 36.1 & \textbf{0.0} & 20.0 & 1.4 & 42.9 & \textbf{0.0} & 49.2 & \textbf{0.4} & 37.0 & 40.4 & 40.2 \\
\grp System    & \sysname{}          & 0.7 & \textbf{78.5} & 3.8 & 65.7 & 0.7 & 76.4 & \textbf{0.0} & \textbf{85.4} & 1.3 & 76.5 & 84.6 & \textbf{83.5} \\
\midrule
\multicolumn{14}{@{}l}{\textit{Llama-3.3-70B}} \\
--        & No defense          & 34.7 & 38.2 & 40.0 & 26.7 & 35.7 & 20.7 & 18.8 & 12.9 & 32.3 & 24.6 & 29.8 & 20.2 \\
\grp Model     & PI detector         & 6.9 & 30.6 & 2.9 & 24.8 & \textbf{0.0} & 5.0 & 6.2 & 7.5 & 4.0 & 17.0 & 20.1 & 19.3 \\
\grp Prompt    & Prompt sandwiching  & 37.5 & \textbf{42.4} & 21.9 & 22.9 & 17.1 & 29.3 & 13.3 & 12.9 & 22.5 & 26.9 & 30.8 & 23.9 \\
          & Spotlighting        & 27.1 & 39.6 & 39.0 & 23.8 & 33.6 & 23.6 & 15.0 & 8.8 & 28.7 & 23.9 & \textbf{38.0} & 27.1 \\
\grp Framework & Progent             & 6.9 & \textbf{42.4} & 10.5 & \textbf{31.4} & 8.6 & 32.9 & 5.8 & 21.2 & 8.0 & 32.0 & 31.8 & 29.3 \\
          & CaMeL               & \textbf{0.0} & 38.2 & \textbf{0.0} & 21.9 & \textbf{0.0} & 0.7 & \textbf{0.0} & \textbf{43.3} & \textbf{0.0} & 26.0 & 26.5 & 26.5 \\
\grp System    & \sysname{}          & \textbf{0.0} & 40.3 & 3.8 & 29.5 & 5.7 & \textbf{38.6} & \textbf{0.0} & 27.5 & 2.4 & \textbf{34.0} & 32.3 & \textbf{31.5} \\
\bottomrule
\end{tabular}
\begin{tablenotes}[flushleft]\footnotesize
\item The \sysname{} oracle and Progent's policy model are GPT-4.1 for every agent. CaMeL runs with its security policies enforced, and its privileged and quarantined models are the agent model of each block.
\end{tablenotes}
\end{threeparttable}
\end{table*}

\begin{table}[t]
\providecommand{\miss}{\textcolor{red}{?}}
\providecommand{\grp}{\rule{0pt}{10.8pt}}%
\centering
\caption{Agent and oracle models varied independently (\%, AgentDojo Banking).}
\label{tab:placement}
\small
\setlength{\tabcolsep}{4pt}
\begin{threeparttable}
\begin{tabular}{@{}llrrr>{\columncolor{nrpbg}[\tabcolsep][0pt]}r@{}}
\toprule
\textbf{Agent} & \textbf{Oracle} & \textbf{ASR}$\downarrow$ & \textbf{U}$\uparrow$ & \textbf{U\textsubscript{b}}$\uparrow$ & \textbf{NRP}$\uparrow$ \\
\midrule
Qwen3.5-35B  & None        & 22.9 & 78.5 & 81.2 & 62.6 \\
             & Qwen3.5-35B & 1.4 & 77.1 & 87.5 & 86.3 \\
             & Qwen3.5-27B & 0.0 & 80.6 & 87.5 & 87.5 \\
             & Qwen3.5-9B  & 1.4 & 79.2 & 93.8 & 92.5 \\
\grp Qwen3.5-27B  & None        & 53.5 & 75.7 & 87.5 & 40.7 \\
             & Qwen3.5-35B & 0.7 & 85.4 & 93.8 & 93.1 \\
             & Qwen3.5-27B & 0.7 & 73.6 & 93.8 & 93.1 \\
             & Qwen3.5-9B  & 0.7 & 77.8 & 81.2 & 80.6 \\
\grp Qwen3.5-9B   & None        & 37.5 & 68.1 & 81.2 & 50.8 \\
             & Qwen3.5-35B & 2.1 & 68.8 & 81.2 & 79.5 \\
             & Qwen3.5-27B & 0.0 & 71.5 & 81.2 & 81.2 \\
             & Qwen3.5-9B  & 0.0 & 61.8 & 68.8 & 68.8 \\
\bottomrule
\end{tabular}
\begin{tablenotes}[flushleft]\footnotesize
\item None: the undefended agent. Qwen3.5-35B is the mixture-of-experts model Qwen3.5-35B-A3B (3B active parameters).
\end{tablenotes}
\end{threeparttable}
\end{table}

\subsection{Per-Suite Results on AgentDojo}
\label{app:results}
This appendix presents a detailed breakdown of AgentDojo results from \cref{sec:eval} by suite. \Cref{tab:main} displays attack success, utility under attack, and benign utility per suite for each defense and all three agent models, using GPT-4.1 as the oracle. The mean columns correspond to \cref{tab:agentdojo-main}.

\subsection{Agent and Oracle Model Pairings}
\label{app:placement}
\Cref{tab:placement} presents the complete results underlying \cref{tab:model-grid}, including attack success, utility under attack, benign utility, and NRP for each agent and oracle model pairing, as well as the undefended agent.
All pairings were evaluated using the AgentDojo Banking suite.

\end{document}